\documentclass{WileyMSP-template}

\usepackage{hyperref}
\usepackage{graphicx}%
\usepackage{amsmath,amssymb,amsfonts}%
\usepackage{xcolor}
\usepackage{import}
\usepackage{xifthen}
\usepackage{pdfpages}
\usepackage{transparent}

\newcommand{\edit}[1]{\textcolor{black}{#1}}

\begin{document}

\pagestyle{fancy}
\rhead{\includegraphics[width=2.5cm]{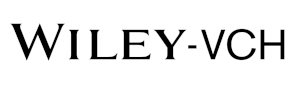}}

\title{Design and Reprogrammability of Zero Modes in 2D Materials from a Single Element}

\maketitle


\author{Daniel Revier}
\author{Molly Carton}
\author{Jeffrey I. Lipton*}


\begin{affiliations}
Daniel Revier\\
Paul G. Allen School of Computer Science and Engineering, University of Washington, Seattle, WA, 98195\\
Email Address: drevier@uw.edu

\medskip

Molly Carton\\
Mechanical Engineering, University of Maryland, College Park, MD, 20742 \\
Email Address: mcarton@umd.edu

\medskip

Jeffrey I. Lipton \\
Mechanical and Industrial Engineering, Northeastern University, Boston, MA, 02115 \\
Email Address: j.lipton@northeastern.edu

\end{affiliations}



\begin{abstract}

Mechanical extremal materials, a class of metamaterials that exist at the bounds of elastic theory, possess the extraordinary capability to engineer any desired elastic behavior by harnessing mechanical zero modes --- deformation modes that demand minimal or\edit{, ideally,} no elastic energy.
However, the potential for arbitrary construction and reprogramming of metamaterials remains largely unrealized, primarily due to significant challenges in qualitatively transforming zero modes within the confines of existing metamaterial design frameworks.
This work presents a method for explicitly defining and in situ reprogramming zero modes of two-dimensional extremal materials by employing straight-line mechanisms (SLMs) and planar symmetry, which prescribe and coordinate the zero modes, respectively.
\edit{We validate the concept experimentally on square‑symmetric lattices and corroborate its generality for hexagonal lattices through finite‑element analysis, together spanning the full theoretical gamut of extremal behaviors.}
The method is used to design, test, and reprogram centimeter-scale isotropic, orthotropic, and chiral extremal materials by reorienting the SLMs in place, enabling these materials to smoothly and reversibly interpolate between extremal modalities (e.g., unimode to bimode), material properties (e.g., negative to positive Poisson's ratios)\edit{, and selectively enable chirality} without changing the metamaterial's global structure.
This methodology provides a straightforward and explicit strategy for the design and tuning of all varieties of two-dimensional extremal materials, enabling 
dynamic mechanical metamaterial construction to completely cover the gamut of elastic properties.

\end{abstract}


\section{Introduction}
\emph{Mechanical extremal materials} \edit{operate at the theoretical limits of linear elasticity and}
are characterized by the number and type of \emph{zero modes}, i.e., deformation modes that cost little or\edit{, ideally,} no elastic energy~\cite{Milton1995-rd, Lubensky2015-iz, Hu2023-md}.
These materials have been developed for applications that include mechanical cloaking~\cite{Buckmann2014-tl, Nassar2019-wy, Xu2020-mc}, acoustic cloaking~\cite{Layman2013-ec, Chen2017-mq}, and out-of-plane shear wave polarizers~\cite{Wei2021-kk}.
Significantly, by coordinating and composing zero material modes, we can theoretically construct any feasible linear elastic material behavior~\cite{Milton1995-rd}.
Although recent advances have brought reprogrammability to extremal materials, allowing them to switch between a preset number and type of zero modes~\cite{Hu2023-md}, the designs are highly specialized, and their zero modes cannot be arbitrarily constructed.
\edit{Crucially, there has not been a demonstration of a single framework to produce a variety of zero modes with different material symmetries, e.g., chiral, orthotropic, or isotropic.}
Without a structured design framework, previous work has resorted to application-specific designs of metamaterials using human effort or costly computational techniques~\cite{Bossart2021-ek, Sigmund2000-gg, Andreassen2014-tc, Wang2014-fk, Sigmund1995-xp, Sigmund1994-qf, Sigmund1994-jx, Cai2022-gm, Yera2020-pd, Hopkins2011-yu, McCarthy2023-ka, Shaw2019-ap}. 
\edit{Thus, a framework that is able to explicitly define, tune, and coordinate the zero modes of a material is a far more comprehensive method to create extremal behavior.}

\edit{Here we show an expressive and reprogrammable method to directly engineer zero modes for two-dimensional (2D) extremal materials using compliant straight-line mechanisms (SLMs) and planar symmetry.}
We base our design on the SLM because it is inherently extremal.
The SLM constrains motion to a single straight-line trajectory~\cite{Kempe1877-gm, Howell2013-yu, Howell2013-ty, Hawks2016-cf, Hubbard2004-md}, and thus operates as an explicitly definable zero mode unit cell.
\edit{We also design our SLMs to be} innately reprogrammable through a rotationally symmetric \edit{construction}, allowing them to pivot in place and modify the zero modes of the metamaterial without altering the global structure.
\edit{To realize the SLMs as a comprehensive metamaterial, the global construction and coordination of the zero modes is performed using planar symmetry which integrates multiple SLMs onto 2D lattices to achieve the desired extremal behaviors.}

This approach achieves centimeter-scale extremal materials of various material symmetries --- isotropic, orthotropic, and chiral. 
It also includes the ability to reprogram these materials in situ to smoothly and reversibly interpolate between different extremal modes (e.g., unimode to bimode) and emergent properties (e.g., negative to positive Poisson's ratio).
Unlike other methods, our \edit{framework} enables \edit{rapid, arbitrary}, continuous, and spatially independent adjustment of engineered zero modes.
Thus, we \edit{are able to} uniquely realize Milton and Cherkaev's concept of \edit{2D} extremal material lamination~\cite{Milton1995-rd} and facilitate dynamic adjustments of material properties on demand \edit{without costly computational methods or lengthy reprogramming procedures.} 
We design, simulate, fabricate, and validate the SLMs and symmetry as a programmable and tunable framework to define the zero modes of extremal materials, \edit{demonstrating how SLM metamaterials can access all possible 2D mechanical extremal modes.}
By examining both the engineering constants (Young's moduli (\(E_1,E_2\)), shear modulus (\(G_{12}\)), Poisson's ratios (\(\nu_{12},\nu_{21}\)), \edit{the normal-shear coupling ratio (\(\eta_{121},\eta_{122}\)}), and the zero modes of the extremal materials' homogenized elastic matrix \(C\), we discover that the orientation of the SLM and the employed symmetry pattern comprehensively determine the emergent properties and extremal behavior.

The core contributions of this paper include: 
(1) the use of SLMs as continuously reprogrammable zero-mode metamaterial cells, 
(2) the use of symmetry to coordinate zero modes and realize extremal types \edit{of varied material symmetries (isotropic, orthotropic, chiral)}, and 
(3) a demonstration of smoothly reprogramming extremal mode and emergent properties of our extremal materials.
Leveraging these insights and contributions, we extend and realize the design space of reprogrammable 2D \edit{extremal} metamaterials and demonstrate their temporal and spatial variability.

\section{Innovation and Methodology}

\begin{figure}[htbp]
    \centering
    \includegraphics[width=0.8\textwidth]{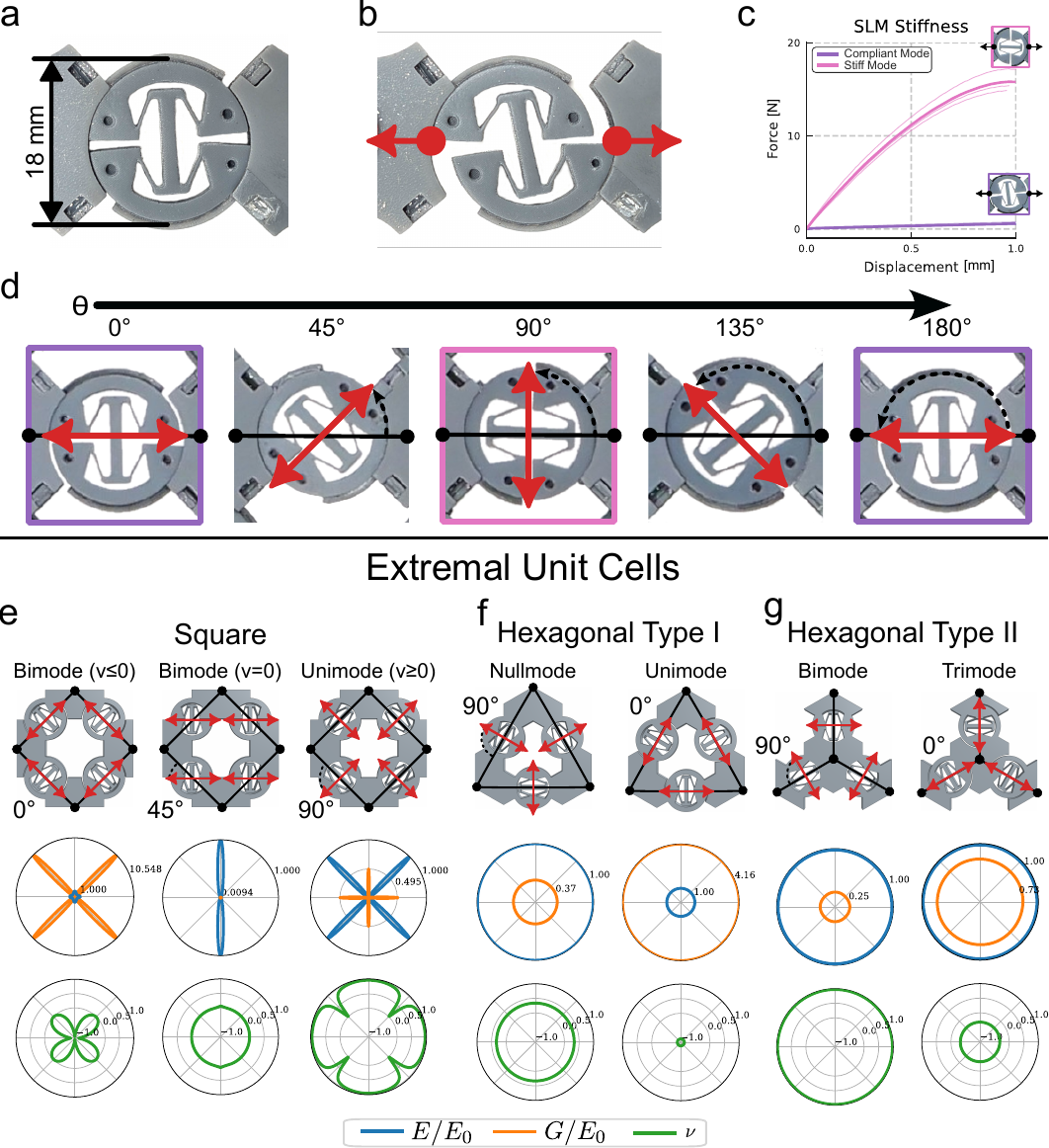}
    \caption{The straight-line mechanism (SLM) enables reprogrammable material properties by modifying the zero modes of extremal materials. The SLM is shown at rest (a) and under horizontal loading (b) \edit{showing linear displacement along the straight-line trajectory when loaded horizontally.} (c) A single fabricated SLM shows an approximate 30-fold reduction in stiffness from rigid loading (pink) to compliant loading (purple). Measured data is shown as lighter lines with averages (n=5) shown as bold lines. (e) The SLM is reprogrammed by pivoting in place and is parameterized by \(\theta\), the angle of the zero mode relative to the lattice connection direction. The SLM can be embedded onto 2D lattices using square symmetry (e) or two different types of hexagonal symmetry (f and g). Each of these configurations (e-g) is shown as CAD unit cell overlaid with its abstract lattice representation and FEA homogenized data shown in the polar plots below. Via in situ reprogramming, these materials can achieve substantially different extremal modes and resulting material properties without affecting the global lattice structure, such as transforming from a negative Poisson's ratio to a positive Poisson's ratio as shown in (e) or predicted (FEA) to support all four types of 2D isotropic extremal materials (f and g). Raw values for moduli are provided in SI Tables S3 and S4.}
    \label{fig:1}
\end{figure}

\subsection{\edit{SLMs are Reprogrammable Unimode Unit Cells}}
SLMs are the foundational element of our extremal materials. 
Among the various SLM designs~\cite{Kempe1877-gm}, we devised a compliant double-sided Robert's mechanism, inspired by prior work on generating a compliant single degree of freedom (DOF) motion~\cite{Hubbard2004-md} (see~\textbf{\autoref{fig:1}}a and b). 
They are compliant exclusively along their straight-line DOF and rigid against orthogonal deformations, with an approximate 30-fold decrease in stiffness between rigid and compliant loading shown in \autoref{fig:1}c. 
SLMs are therefore inherently extremal, \edit{specifically unimodal, possessing one intrinsic zero mode.}
\edit{As shown by~\cite{Milton1995-rd} unimodes are sufficient to form the basis for all other extremal materials, thus our SLM-based framework is well-suited for extremal material construction.}

A key innovation of our approach is the SLM's rotationally symmetric design, which allows in situ reorientation of the compliant direction (see~\autoref{fig:1}d); this tunes the zero modes of our extremal materials.
\edit{The} idealized SLM compliance \edit{can be expressed} as the deformation gradient \(F\) of the SLM cell base and the subsequent linear strain \(\varepsilon = \frac{1}{2}(F+F^T)-I\)
\begin{equation}
    F
    =
    \begin{bmatrix}
        1 + \alpha \cos{\theta} & 0 \\
        \alpha \sin{\theta} & 1
    \end{bmatrix}
    \label{eq:F}
\end{equation}
and
\begin{equation}
    \varepsilon = 
        \frac{\alpha}{2}
    \begin{bmatrix}
        2\cos\theta & \sin\theta \\
        \sin\theta & 0
    \end{bmatrix}
    \label{eq:strain}
\end{equation}
where \(\alpha\) is the non-dimensional length scale of the SLM deformation and \(\theta\in[0\degree,180\degree)\) is the angle of the SLM DOF relative to the connective direction, shown in~\autoref{fig:1}d.
The linear strain presented here is equivalent to the unimode strain expressed in Milton's and Cherkaev's foundational work~\cite{Milton1995-rd}.
Thus, our SLM fulfills the original criterion for extremal material construction from unimode and can simultaneously adapt its extremal material response --- from shear to normal deformation and back --- through simple rotational adjustments.
These unique characteristics enable in situ tuning of the mechanical behavior without requiring global structural change or lengthy reprogramming processes.

\subsection{\edit{Zero Modes are Composable Using Planar Symmetry}}
Na\"ively tiling unimodal SLMs can only result in unimodal materials with compliant strains described by~\autoref{eq:strain}. 
\edit{Thus,} introducing more \edit{DOF-wise} independent SLMs is necessary to achieve a broader range of behaviors; however, each \edit{adjoined} SLM introduces \edit{constraints, (i.e., rigidity)}
along with its DOF \edit{as expressed by the zero value in~\autoref{eq:strain}}. 
\edit{Crucially, the DOFs and constraints} must be coordinated to ensure that they do not interfere 
and restrict deformation, which requires simultaneously satisfying each individual SLM's deformation gradient.
See SI \S 1 for more details.

We use \edit{2D} planar symmetries, called the \emph{wallpaper groups}, to introduce multiple SLMs and coordinate their zero modes.
Wallpaper groups ensure complete tiling of the plane for metamaterial construction~\cite{Conway2016-rp}.
They also effectively arrange each individual SLM zero mode to avoid the combinatorial complexity seen in other efforts to design zero mode materials~\cite{Bossart2021-ek}.
This setup creates a two-dimensional material where the number and characteristics of zero modes are determined by the SLM orientation and \edit{applied} symmetry.
\edit{Due to its uniform structure,} our reprogrammable SLM 
fits seamlessly into regular square and hexagonal lattices.
\edit{Significantly,} in contrast to other reprogrammable zero mode materials~\cite{Hu2023-md}, the overall global lattice structure is preserved regardless of the microstructure \edit{configuration (i.e., SLM orientation), demonstrated in~\autoref{fig:1}e-g.}
This allows the lattice to act as a scaffold, where the nodes serve as connection points for the SLMs positioned along the edges.
The loading of the structure is transferred through the lattice connections, but deformations are allowed only in line with the SLM orientations.
This results in 
\edit{a range of emergent behaviors based on} 
a single lattice, which are coordinated by the orientation of zero modes relative to the lattice lines. 

\subsection{Analyzing Zero Modes}
\edit{First, we developed an analytical model for the compliance matrix \(S=C^{-1}\) using the idealized deformation gradient and linear strains (see SI \S 2 for derivation).
We then simulated the materials using finite element analysis (FEA) computational homogenization via ANSYS Material Designer (AMD) to extract an effective metamaterial linear elastic response across \(\theta\).
Due to limitations of AMD, only orthogonal tilings are possible, precluding the ability to model the 632 material in the range of \(\theta\in(0\degree,90\degree)\).
This allows us to compare theoretical models, fit data, and simulation for the full range of 2*22 and 442, but only the orthogonal modes of 632.
}

\edit{
The square-symmetry materials (2*22 and 442) were tested in different \(\theta\) configurations using a \(2 \times 10\) array under tensile and shear loading, with optical tracking to extract the engineering constants of the array (Young's moduli \(E_1,E_2\), shear modulus \(G_{12}\), Poisson's ratios \(\nu_{ji}=\varepsilon_i/\varepsilon_j\), and normal-shear coupling \(\eta_{12k}=\gamma_{12}/\varepsilon_k\)). 
All mechanical tests were conducted with \(<2\%\) global strain where linear elasticity is a reasonable approximation for extracting effective properties.
The engineering constants were then used to recreate the full compliance matrix \(S=C^{-1}\) for comparison with FEA results.
}

\edit{
In order to directly compare between the FEA and measured results, we fit both sets of data to the analytical model using linear least-square regression with non-linear constraints (see SI \S 2).
This results in a more numerically robust and higher-order model that characterizes the entire system across \(\theta\).
In turn, we were able to investigate how the models behave holistically rather than on a point-wise basis for each \(\theta\) value, leading to more effective comparisons between FEA modeled and fabricated/simulated tests.
We then analyzed the zero modes of these fitted models, unconvering relationships that would otherwise be obscured by numerical and measurement noise.
}

\section{Results \edit{\& Discussion}}
\subsection{SLM Orientation and Symmetry Provide Different Emergent Properties}
We observe a variety of emergent and extremal properties by constructing materials using wallpaper groups 2*22, 632, and 442 expressed here in the orbifold notation~\cite{Conway2016-rp} (cmm, p6, p4 in crystallographic notation) and shown in \autoref{fig:1}\edit{e-g}.
In all cases, altering the orientation of the SLM within a single lattice type maintains the global structure, but leads to the development of distinct extremal materials.

\autoref{fig:1}e shows \edit{the topological layout and FEA simulated data for} three variations of the 2*22 symmetry for \(\theta \in \{0\degree,45\degree,90\degree\}\), which create three distinct orthotropic extremal materials: an auxetic bimode, a zero Poisson's ratio bimode, and a positive Poisson's ratio unimode.
In each case, the behavior of the material \edit{is described by} 
\autoref{eq:strain} 
\edit{which prescribes the directions of compliance relative to the lattice lines.}

\edit{
The angles of \autoref{fig:1}e show a wide range of simulated FEA properties.
For \(\theta=0\degree\) the zero modes of a material will lie diagonally along the lattice lines, without any shear compliance.
With SLMs along each diagonal of the lattice, this material is a bimode, compliant with extension along each of the diagonals independently.
This is demonstrated in the moduli plot along the diagonal, where a low Young's modulus and a high shear modulus confirm the expected allowable deformations.
Additionally, there is a zero Poisson's ratio along those diagonals with the DOFs, but the diverging nature of the DOFs causes a negative Poisson's ratio (auxetic) along the \edit{principal} horizontal and vertical directions.
The 2*22 \(\theta=45\degree\) structure is also a bimodal material, allowing independent horizontal and shear deformations due to the alignment of zero modes in the horizontal direction.
This results in a low Young's modulus horizontally, a high Young's modulus vertically, a low shear modulus in the plane, and a zero Poisson's ratio in the plane.
Finally, the 2*22 structure with \(\theta=90\degree\) exhibits unimodal properties.
Complementary to the \(\theta=0\degree\) case, here only the off diagonal elements of \autoref{eq:strain} are non-zero, meaning that the material does not extend and can only shear relative to the lattice lines.
This is seen with the low shear modulus along the diagonals.
Furthermore, in contrast to \(\theta = 0\degree\), the DOFs are now aligned to produce a positive Poisson's ratio in the horizontal and vertical directions.
Taken together, these three orientations prove that a single lattice can be dialed from stretch, to stretch-shear, and then shear‑only behavior simply by rotating the SLMs, demonstrating full, on‑demand control over its deformation modes.
}
\newline
\edit{Our analytical and FEA models indicate that hexagonal 632 symmetry can realize all four isotropic extremal types --- null-, uni-, bi-, and trimode --- via two tiling variants (\autoref{fig:1}f and g).
Strikingly, the hexagonal lattices only need two orthogonal orientations (\(\theta=0\degree\) and 90\degree{}) to span this full isotropic extremal set.
In this case the Type I tiling produces both nullmode and unimode materials (90\degree{} and 0\degree{} respectively), while Type II tiling produces bimode and trimode materials (90\degree{} and 0\degree{} respectively).
} 

\edit{
The Type I construction uses an equilateral triangle cell layout, which adds a \edit{geometric} constraint that all sides must be equal upon deformation (see SI \S 1).
The combination of this constraint and zero modes oriented perpendicular to the lattice lines (\(\theta=90\degree\)) results in a material that resists extension or shear deformation, making it a nullmode material.
While nullmode materials are often considered trivial, our SLM-based design produces a non-trivial internal rotational motion due to the coordinated DOFs of its zero modes. 
Notably, this rotational DOF is fully described by the SLM deformation gradient (\autoref{eq:F}), but is not captured in the linear elastic theory (\autoref{eq:strain}).
}

The Type I material with SLMs oriented along the lattice lines (\(\theta=0\degree{}\)) \edit{numerically demonstrates} an isotropic unimode.
The zero modes align with the lattice lines meaning that only expansion and contraction along the lattice dimensions are allowed.
This produces a material that compresses and expands easily without shear, i.e., has a low bulk modulus and high shear modulus, a key feature of isotropic unimodes.
Consequently, this also results in an \edit{FEA simulated} negative Poisson's ratio of \(\nu \approx -0.88\) which approaches the theoretical limit of \edit{$-1$ for isotropic 2D elastic materials}.

\edit{Type II tiling covers the isotropic bimode-trimode materials.}
Similar to the nullmode material, the bimode material has DOFs perpendicular to the lattice \((\theta=90\degree{})\); however, Type II tiling does not have the additional equilateral triangle constraint from Type I.
\edit{Because of this, the bimode material is able to shear along the lattice lines, producing the opposite effect of the Type I unimode, that is a material with a high bulk modulus and low shear modulus.}
This bimode material is the 2D analog of the 3D pentamode material~\cite{Milton1995-rd, Kadic2012-qy}, operating as a ``metafluid'', resisting bulk compression, but compliant to shear or ``flow.''
\edit{Therefore as expected, this material approaches the theoretical limit of +1 for isotropic 2D elastic materials with a simulated value of \(\nu=0.97\).}

Finally, the trimode material, like the nullmode, is normally considered trivial as it is completely compliant; however, our SLM trimode material has DOFs along the lattice lines (\(\theta=0\degree{}\)).
As with the previous materials with DOFs along the lattice, this trimode material 
has a negative Poisson's ratio, \edit{a feature} that has not been previously seen in other compliant extremal materials.
The sum total of our isotropic materials is that we have shown a method of constructing all four types of isotropic extremal materials using SLMs.

\subsection{The Property Space is Interpolative}

\begin{figure}[htbp]
    \centering
    \includegraphics{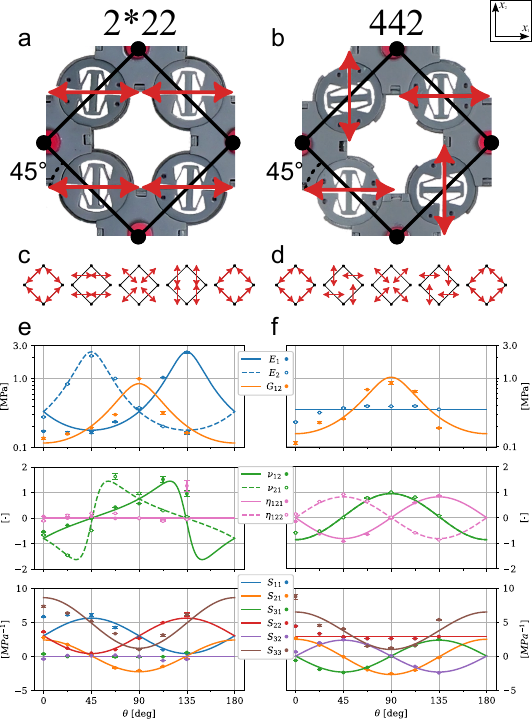}
    \caption{\edit{Fabricated 2*22 (a) and 442 (b) square symmetry materials shown in close-up at a \(\theta=45\degree\) configuration. The illustrations below each image (c and d) show the DOF alignment for each symmetry as a function of \(\theta\) and correspond to the major x-axis grid lines in the plots below. The data shown in (e) and (f) show experimental data (dots) and corresponding analytical fits (solid/dashed lines). From top to bottom: the directional Young's and shear moduli ($E_1,E_2$ and $G_{12}$); the Poisson's and normal-shear coupling ratios ($\nu_{12},\nu_{21}$ and $\eta_{121},\eta_{122}$); and finally the individual $S$ matrix components which were used to fit the model directly. Markers show the median of five repeats, n=5; vertical bars denote interquartile range. Data was not gathered for 157.5\degree{} due to limitations of the dovetail construction. Similarly, data was not gathered for 180\degree{} due to it having an identical construction to 0\degree{} and so was considered redundant.}}
    \label{fig:four-fold}
\end{figure}

In Figure~\ref{fig:1} we see discrete states of the lattice; however, unlike previous work~\cite{Kadic2012-qy, Hu2023-md, Wei2021-kk}, our extremal materials are continuously reprogrammable. 
By rotating the SLMs we produce smooth variations in their properties. 
This produces an interpolative property space parameterized by the continuous variable \(\theta \in [0,180\degree{})\).
This variability arises from both the lattice symmetry and the SLM orientation, each independently influencing the material's properties.
\edit{Thus, by} using a single lattice we can enable or disable properties, like chirality or auxetics, through reprogramming of the SLM orientations.


The two materials shown in~\autoref{fig:four-fold} \edit{demonstrate} how 
the same square lattice can produce distinct 
\edit{interpolative material property characteristics}
by enforcing different symmetry types (mirror vs. rotational) on the SLMs.
\edit{
\autoref{fig:four-fold}a and b show the two square symmetry patterns, 2*22 and 442, in a \(\theta=45\degree{}\) configuration.
Shown directly below (\autoref{fig:four-fold}c and d) is an illustration for how the DOFs differ due to mirror vs. rotational symmetry as \(\theta\) varies is shown directly, which correspond to the x-axis gridlines of the below plots.
Finally, we show the measured data (dots) and fit model data (solid/dashed lines) in \autoref{fig:four-fold}e and f.
The measured data was captured every 22.5\degree{} while the analytical model is fit to these points but interpolated along the full span of \(\theta\).
Each \(\theta\) configuration was tested with five physical replicates with median values reported and vertical bars representing the interquartile range (25th-75th percentiles).
Data is not shown for 157.5\textdegree{} or 180\textdegree{}. 
Data was not captured for 157.5\textdegree{} due to limitations of the dovetail mechanism, while data for 180\textdegree{} was not captured due to being identical by construction to the 0\textdegree{} configuration and so was considered redundant in nature. 
Regardless, the physical performance is expected to follow the relationships driven by the symmetry of the material, predicted by the analytical model, and demonstrated numerically with the FEA data (see SI Figure S3 for FEA model fit).
}

\edit{
For the 2*22 material in~\autoref{fig:four-fold}e, the Young's modulus varies from low to high depending on the alignment of the SLMs, low when all the DOFs are aligned with a principal direction and high in the orthogonal direction, while the shear modulus peaks simply once for \(\theta=90\degree{}\).
Notably, where the Young's moduli peak (45\degree{} \((E_2)\) and 135\degree{} \((E_1)\)) the corresponding Poisson's ratio has an asymptotic behavior, due to the SLM rigid directions all being aligned in the respective principal direction, \(x_1\) or \(x_2\).
While the shear modulus demonstrates a single peak at 90\degree{}, the normal-shear coupling is minimal by the mirror symmetric nature of the 2*22 pattern, with the exception at 135\degree{}, which is an artifact of testing at the asymptotic point highlighted earlier for Poisson's ratio.
Significantly, the moduli can be varied by an order of magnitude and along with that we see the Poisson's ratio vary from negative (auxetic) to positive.
}

\edit{
In contrast to the 2*22 material, the 442 Young's modulus remains relatively flat (\autoref{fig:four-fold}f).
This is expected because all four DOFs of the unit cell never fully align and, by rotating together, lead to a constant level of stiffness as a function of \(\theta\).
The shear modulus, however, follows the same trend as the 2*22 material and also varies about one order of magnitude.
Here the Poisson's ratio does not exhibit the asymptotic behavior, instead producing a cosine wave, while the normal-shear couplings produce sine waves of opposite signs (\(\eta_{121}=-\eta_{122}\)).
This normal-shear coupling is expected due to the rotational symmetry of the material which, lacking mirror symmetry, allows for non-zero \(S_{13}\) and \(S_{23}\) values, a feature absent with the 2*22 material.
Importantly, we see the chirality of the material changes as a function of \(\theta\) with varying \(\eta_{121}\) from $-1$ to $+1$. This allows us to turn on and off the coupling as well as change the handedness of the material.
}

\edit{
The bottom plot of~\autoref{fig:four-fold}e and f show the measured data and corresponding analytical model of the \(S\) matrix.
The specific coefficients of the analytical model are then optimized to minimize the relative error across all \(S\) values equally and are provided in Table S2 (SI).
Overall, there is good agreement between the analytical model and measured data with several key idealizations were made in generating the models.
First for the 2*22 material, the values of \(S_{31}\) and \(S_{32}\) were assumed to be zero, due to the natural symmetry of the material.
This is demonstrated in the analytically derived model and FEA data.
While the data does not reflect zero values, it is still relatively small when compared to other values of \(S\), most likely arising from noise in the motion capture system.
For the 442 material, the values of \(S_{11}\) and \(S_{22}\) were assumed to be constant, again supported by the analytical model and corroborated by the FEA simulated data.
These simplifying assumptions enable the model to capture the dominant trends in material behavior across all \(\theta\), while smoothing out local noise and inconsistencies in the measured data—resulting in a compact, robust representation that reflects both the symmetry-driven physics of the system and the empirical observations.
}

\subsection{The Extremal Materials Span the Eigenvalue Gamut}
\begin{figure}[htbp]
    \centering
    \includegraphics[width=\textwidth]{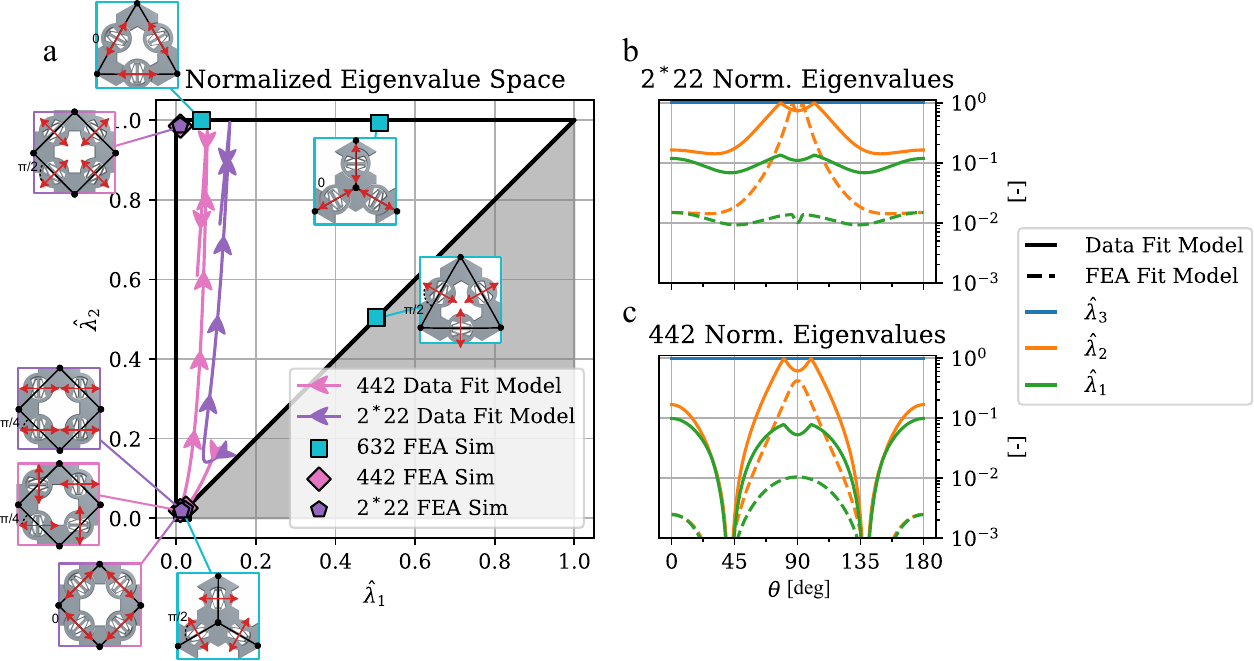}
    \caption{(a) Normalized FEA simulated eigenvalues of the $2^*22$, $442$ and $632$ materials plotted as $\hat{\lambda}_2$ vs. $\hat{\lambda}_1$. Additionally, the transition from bimode to unimode of the fitted data models for the $2^*22$ (purple) and $442$ (pink) are presented for $\theta=[0\degree,90\degree]$. The arrowheads indicate the direction of the trajectory as $\theta$ transitions from 0\degree{} to 90\degree{}. (b and c) The normalized eigenvalues from the experimental data and FEA data analytical fits for the $2^*22$ and $442$ materials respectively. The line style indicates the measured data and FEA models (solid and dashed) and the line color indicates the normalized eigenvalue being plotted for a given model (\(\hat{\lambda}_3\) is blue, \(\hat{\lambda}_2\) is orange, \(\hat{\lambda}_1\) is green). Absolute and relative eigenvalues for all FEA points are tabulated in Table S4 (SI).}
    \label{fig:gamut}
\end{figure}

\edit{
To analyze the transitions of SLM lattices, we define the ``extremal material gamut'' (\autoref{fig:gamut}) which provides a stiffness- and coordinate-invariant space to compare materials.
Here we use normalized eigenvalues of \(C\) sorted smallest to largest, constraining the values within the upper-left triangle of the unit square.
Specifically, if the eigenvalues of the \(3\times3\) \(C\) are \(0<\lambda_1 \le \lambda_2 \le \lambda_3\) then the normalized eigenvalues are \(\hat{\lambda}_i=\lambda_i/\lambda_3\) for \(i=1,2,3\).
Because \(\hat{\lambda}_3=1\) by definition, we plot the relationships in a 2D space \(\hat{\lambda}_2\) vs. \(\hat{\lambda}_1\).
In this space, the diagonal represents \(\lambda_2=\lambda_1\) while the top line represents \(\lambda_2=\lambda_3\).
The remaining left-most boundary forms the extremal boundary, where there is always one small relative eigenvalue (\(\lambda_1 \ll \lambda_3\)) and defines the ideal unimode and bimode materials at points \((0,1)\) and \((0,0)\) respectively.
Importantly, because we normalize each material configuration at \(\theta\) to \emph{its own} largest eigenvalue we can compare unimode and bimode materials directly in terms of their extremal characteristics independent of their inherent stiffness.
The tradeoff for this normalization is that it obscures materials with three small, but relatively close, eigenvalues, such as the 632 trimode discussed below.
}

\edit{
As shown in~\autoref{fig:four-fold}, just as the material properties smoothly vary over \(\theta\), so do the eigenvalues.
This is seen by the purple and pink trajectories for the measured-data model in \autoref{fig:gamut}a.
The significance of the trajectories is two-fold: 1) both material sets (2*22 and 442) maintain relative extremality and 2) the extremal space is continuously variable.
Here, the trajectories span \(\theta=0\degree{}\) to \(\theta=90\degree{}\) and follow the flow of the arrows. 
\autoref{fig:gamut}b and c show the normalized eigenvalues of these trajectories as a function of \(\theta\).
In each case the analytical 2*22 and 442 materials consistently have one small eigenvalue \((\hat{\lambda}_1 \lesssim 10^{-1})\) while \(\hat{\lambda}_2\) varies from low to high, nearly equal with \(\hat{\lambda}_3\) as \(\theta\rightarrow90\degree{}\).
This means the trajectories in the gamut (\autoref{fig:gamut}a) show the square-symmetry materials transitioning from bimode to unimode by tracking along the far-left extremal boundary.
The back-tracking seen toward the end of these trajectories, while remaining extremal, indicates a reordering of the eigenvalues relative to the eigenvectors, which is an artifact of the eigenvalue normalization process.
}

\edit{
Beyond the measured-data model, we fit the FEA simulated data in the same manner for direct comparison.
These are represented as the scatter data points on~\autoref{fig:gamut}a and the dashed lines in~\autoref{fig:gamut}b and c.
Here we see all the examples from~\autoref{fig:1}e-g, now shown in an extremal landscape. 
The 2*22 and 442 \(\theta=0\degree{}\) and \(\theta=45\degree{}\) materials are all bimodal, existing at the bottom left corner, noting that the 2*22 and 442 \(\theta=0\degree{}\) materials are identical in DOFs and so overlap here.
The 2*22 and 442 \(\theta=90\degree{}\) material (again identical in DOFs) is shown at the top-left corner clearly exhibiting unimodal characteristics.
Significantly, this demonstrates that the square-symmetric materials are capable of maintaining extremality while transitioning from bimode to unimode.
}

\edit{
The 632 materials from~\autoref{fig:1}f and g are all shown.
The null mode material (Type I tiling, \(\theta=0\degree{}\)) lies plainly in the middle of the space with no small eigenvalues approximately at (0.5, 0.5).
For the extremal materials, the isotropic unimode (Type I tiling, \(\theta=0\degree{}\) lies at the to left corner, while the bimode (Type II tiling, \(\theta=90\degree{}\)) lies at the bottom left corner, matching their desired extremal behavior.
Notably, the trimode material (Type II tiling, \(\theta=0\degree{}\)) does not lie on an extremal boundary.
While counterintuitive, this is an artifact of the normalization process.
This gamut obscures materials where all the eigenvalues are small relative to the base material, as in the 632 trimode case, which lies near \((0.5,1)\) despite have three simulated small eigenvalues, as shown Table S3 in the SI.
Regardless, the fact that all the unimode and bimode materials lie in their respective domains, and that the trajectories of the square-symmetric materials track along the extremal boundary, confirms the desired outcome of 1) programming all types of extremal behavior and 2) the ability to smoothly interpolate between extremal points.
}


\section{Spatially Varying Orientations Make Programmable Composites}
\begin{figure}
    \centering
    \includegraphics[width=0.8\textwidth]{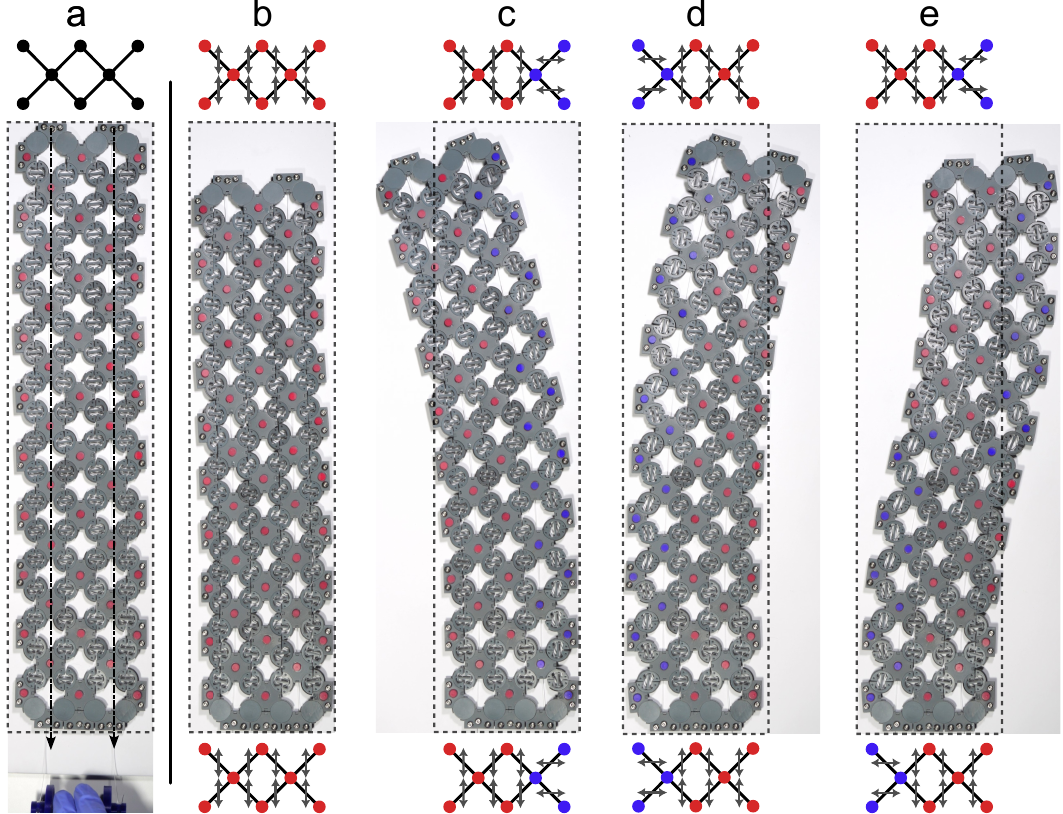}
    \caption{A single, cable-driven array can be spatially reprogrammed to provide unique actuation modes. (a) depicts an illustration of the lattice structure in the undeformed state while (b)-(e) show deformed states. Markers are physically red; in each panel we recolor subsets to match the DOF groupings shown in the lattice illustrations. SLM orientations are depicted by gray arrows. (b) has all the SLMs oriented vertically and compresses with $\nu\approx0$ under tension. (c) and (d) have one column of horizontally oriented SLMs, right and left respectively causing the lateral bends. (e) uses different SLM orientations on the top and bottom halves of the array to create the composite S-bend.}
    \label{fig:4}
\end{figure}

\edit{To demonstrate the utility of the reprogrammable SLM framework we implemented a 2D column with spatially tailored material properties.}
By globally coordinating the changes in the angle \(\theta\) of the SLMs, we control the properties of the entire lattice\edit{; however, the} angle \(\theta\) also provides us with a field \edit{to spatially vary the local} properties of the lattice.
We take advantage of this tunability by reprogramming local stiffness in a finite array to provide different modes of actuation, as shown in \textbf{\autoref{fig:4}}.

We \edit{utilized our} \(2 \times 10\) array \edit{from measurement characterization} to spatially compose different stiffnesses by varying SLM alignment throughout the structure.
To load the array, we affixed two cables at the top, routed them downward through the body, and connected them to the array throughout the length to prevent buckling (\autoref{fig:4}a). 

When all SLMs are aligned with the loading direction (\autoref{fig:4}b), the material compresses in the vertical direction. 
This lattice has the properties of the 2*22 material previously described in \autoref{fig:four-fold}a for \(\theta = 135\degree{}\)\edit{; namely low \(E_2\) and \(\nu_{21}\approx0\)).}

In contrast, reprogramming one column of SLMs perpendicularly partially emulates the 2*22 material for \(\theta = 45\degree{}\) along that vertical band (i.e., high \(E_2\)).
The outcome of the striping process is a material that exhibits low stiffness on one side and high stiffness on the other in relation to the loading direction. 
\edit{Notably, because both of these configurations have low and equal shear modulus (seen in~\autoref{fig:four-fold}e) the material is allowed to bend easily without significant shear stresses.}
This results in lateral bends as only one side of the array \edit{compresses}, and by selecting the right or left most column for stiffening, we can control the direction in which the array bends (\autoref{fig:4}c and d).

\edit{Leveraging the spatial reprogrammability even further,} the design allows for \edit{more sophisticated behaviors}, as illustrated by the S-bend curve pattern in \autoref{fig:4}e.
We take the lattice in \autoref{fig:4}a, split it vertically into two halves, and program the upper half like \autoref{fig:4}c and the bottom like \autoref{fig:4}d. 
This leads to an S-shaped curve when compressed \edit{with complementary bending}. 
The adaptability shows that our SLM-based materials enable the development of spatially programmable mechanical metamaterials within a single lattice \edit{and furthermore is not restricted to simple configurations.}
\edit{Significantly, a myriad of behaviors including auxetics and chirality are technically achievable as long as the DOFs and constraints are resolved without frustration.}

\section{Conclusion}
\edit{
In this study, we introduced a novel, reprogrammable approach to engineering 2D extremal materials using straight-line mechanisms and planar symmetry. 
SLMs provide a geometric basis for explicitly programming the zero modes of extremal materials, enabling experimental construction for square lattices and numerical validation for hexagonal lattices across nullmode through trimode responses, dictated by SLM orientation ($\theta$), lattice type (square vs. hexagonal), and orientation symmetry (rotational vs. mirror). 
By controlling the SLM orientation, we can spatially vary elastic properties across a given lattice, allowing selective tuning of Poisson's ratio, bulk-to-shear stiffness ratio, and chirality within a unified framework. 
We demonstrated the breadth of this approach by simulating and experimentally validating the square-symmetric unimode and bimode emergent material properties and extremal behaviors a stiffness- and coordinate-invariant extremal material gamut.
We also demonstrate numerically how two types of hexagonal tilings can produce all four types of 2D isotropic extremal materials. 
This work not only realizes distinct extremal states but also demonstrates smooth, continuous transitions between them, representing the first such achievement in extremal materials.
}

\edit{
While our results demonstrate the ability of a single reconfigurable unit to access all four extremal deformation classes, practical limitations remain. 
Our simulations were primarily restricted to rectangular tilings due to ANSYS limitations with non-rectilinear periodic boundary conditions, leading us to fabricate only square-symmetric specimens. 
Future work will experimentally confirm the predicted continuous behavior of hexagonal specimens. 
Although experiments were conducted within the linear elastic regime (under 2\% global strain), local large strains, geometric nonlinearities, buckling, or friction were not explicitly modeled and may influence behavior at larger deformations. 
This reliance on standard linear elasticity also meant we could not fully validate intriguing degrees of freedom, such as the internal rotational motion observed rotationally symmetric materials. 
These uncaptured behaviors suggest a promising direction for future theoretical and experimental work exploring connections to micropolar/Cosserat elasticity and their potential for phenomena like mechanical cloaking~\cite{Nassar2019-wy, Xu2020-mc}.
Finally, long-term durability and fatigue resistance of compliant SLM joints were not characterized. 
Future efforts will address material fatigue, creep, and statistical variation across printed samples through higher-throughput testing and extended mechanical cycling to validate performance in real-world scenarios.
}

\edit{
This work significantly expands the design space for mechanical metamaterials, enabling geometrically intuitive, on-demand, dynamic adjustments of mechanical properties without structural alterations or costly computational methods. 
The continuous reprogrammability and arbitrary construction of zero modes unlocked by our SLM-based framework offer unprecedented control over material behavior. 
This paves the way for practical applications demanding adaptive mechanical responses, from next-generation soft robotics and underwater acoustics to advanced wearable technologies, fundamentally shifting how we design and utilize materials with tailored mechanical properties.
}

\section{Experimental Section}
\threesubsection{Fabrication} 
All SLM samples were fabricated using a Carbon 3D M1 printer with UMA-90 material \edit{using standard print settings.}
The SLMs were printed in plane with the print bed and with a Carbon 3D film release coating applied to the print bed. 
Post-processing followed the manufacturer's recommendations.

\threesubsection{Simulation} 
Computational homogenization was performed using ANSYS Material Designer.
Representative volume elements (RVEs) were created in CAD software and simulated with periodic boundary conditions in 3D to extract the full \(6 \times 6\) 3D elasticity matrix.
This matrix was then reduced to the \(3 \times 3\) elasticity matrix for 2D plane stress materials. 
In this work, all analysis of simulated data was done on the reduced 2D matrix. 

RVEs for 2*22 and 442 materials were generated to simulate data for \(\theta \in [0\degree, 180\degree)\) in 5\degree{} steps, using symmetry to minimize the required number of simulations.
632 models were simulated with only \(\theta=0\degree\) and 90\degree{} due to ANSYS Material Designer's limitations of (1) rectilinear periodic boundary conditions and (2) the entire simulated body needing to be one continuous object.

\threesubsection{Test and Characterization} 
A \(2 \times 10\) square-lattice array was fabricated to validate and characterize the 2*22 and 442 materials. 
The design incorporates snap-fit connectors and a rotational dovetail, facilitating reconfigurable and easy assembly. 
This modular construction allows for the easy replacement of parts because the SLMs are not rigidly connected, instead relying on friction and other forces to maintain contact. 
A ratcheting scheme aligns and secures the SLMs during testing.
See SI \S 4 and Figures S3 and S4 for more details.

The array was tested on a universal testing machine (UTM) (Instron 5965) in a quasi-static manner.
For each rotation angle five tests were performed; we report the median value, with error bars spanning the interquartile range (25th-75th percentiles).
Two symmetry patterns (2*22 and 442) were tested, in two different loading conditions (tensile and shear), with approximately five tests per \(\theta\) value for \(\theta \in {0, 22.5^\circ, 45^\circ, 67.6^\circ, 90^\circ, 112.5^\circ, 135^\circ}\).
Measured data could not be captured for \(\theta=157.5\degree\) due to the dovetail mechanisms construction and data at \(\theta=180\degree\) is identical to 0\degree{} by construction, thus this data is not included in the measured data.

The force vs. displacement curves and videos of the array deforming array were recorded for testing at various \(\theta\) configurations.
The global force was used for modulus characterization (\(E_1,E_2,G_{12}\)), while the middle third of the optical trackers were used to capture deformation behavior and characterize strain coupling ratios (\(\nu_{12},\nu_{21},\eta_{121},\eta_{122}\)).
The middle third of the trackers were used to avoid clamped boundary effects at the top and bottom end of the array.

\medskip
\textbf{Supporting Information} \par 
Supporting Information is available from the Wiley Online Library or from the author.

\medskip
\textbf{Data Availability Statement} \par
All data will be available through the Northeastern University Data Repository Service (DRS).

\medskip
\textbf{Funding Statement} \par
The authors would like to acknowledge that this work was made possible by National Science Foundation Grant Nos. 2212049 and 2035717, A gift from the Ford Motor company, and by the  Murdock Charitable Trust through Grant No. 201913596. 

\medskip
\textbf{Conflict of Interest Disclosure} \par
The authors declare no conflict of interests.

\medskip
\textbf{Acknowledgments} \par 
The authors would like to acknowledge Lucas Meza for his advice and guidance as well as Tom Zimet, Aditya Patil, Sam Moorhouse, Jake Simeroth, Almond Lau, and Marco Santonil for their assistance in modeling and fabrication.

\medskip

%
\bibliographystyle{MSP}
\bibliography{bib.bib}




\end{document}


\pagestyle{fancy}
\rhead{\includegraphics[width=2.5cm]{vch-logo.png}}

\title{Design and Reprogrammability of Zero Modes in 2D Materials from a Single Element: Supporting Information}

\maketitle


\author{Daniel Revier}
\author{Molly Carton}
\author{Jeffrey I. Lipton*}

\begin{affiliations}
Daniel Revier\\
Paul G. Allen School of Computer Science and Engineering, University of Washington, Seattle, WA, 98195\\
Email Address: drevier@uw.edu

\medskip

Molly Carton\\
Mechanical Engineering, University of Maryland, College Park, MD, 20742 \\
Email Address: mcarton@umd.edu

\medskip

Jeffrey I. Lipton \\
Mechanical and Industrial Engineering, Northeastern University, Boston, MA, 02115 \\
Email Address: j.lipton@northeastern.edu

\end{affiliations}







\section{Straight Line Mechanism Deformation Gradients and Strains}

\subsection*{Single SLMs}
We analyze the behavior of the SLM extremal materials in two ways: through the deformation gradient and through the linear strain.
We first express the idealized SLM compliance as the deformation gradient \(\mathbf{F}\) of the SLM cell base as seen in the main text Figure 1a
\begin{equation}
    \mathbf{F}
    =
    \begin{bmatrix}
        1 + a \cos{\theta} & 0 \\
        a \sin{\theta} & 1
    \end{bmatrix}
    \label{eq:F},
\end{equation}
where \(a\) is the length scale of the SLM deformation and \(\theta\in[0,\pi)\) is the angle of the SLM DOF (red arrow) relative to the connective direction (black solid line).
We can then express the SLM's zero mode as a linear strain
\begin{equation}
    \varepsilon = \frac{1}{2}(\mathbf{F}+\mathbf{F}^T) - \mathbf{I}
    =
    \frac{a}{2}
    \begin{bmatrix}
        2\cos{\theta} & \sin{\theta} \\
        \sin{\theta} & 0
    \end{bmatrix}
    \label{eq:strain}
\end{equation}
which is equivalent to the unimode material described in~\cite{Milton1995-rd} Equation 2.6. 

Notably, the SLM and its deformation gradient exhibit rotational symmetry (C2 point group), whereas the strain shows mirror symmetry (D2 point group) for \(\theta\) values of 0 and 90 degrees.
Therefore, even if the symmetry pattern is solely rotationally symmetric, such as 632, the material properties can display characteristics requiring mirror symmetry, like isotropy.

\subsection*{Lattice and Deformation Vector Construction}

\begin{figure}
    \centering
    \includegraphics{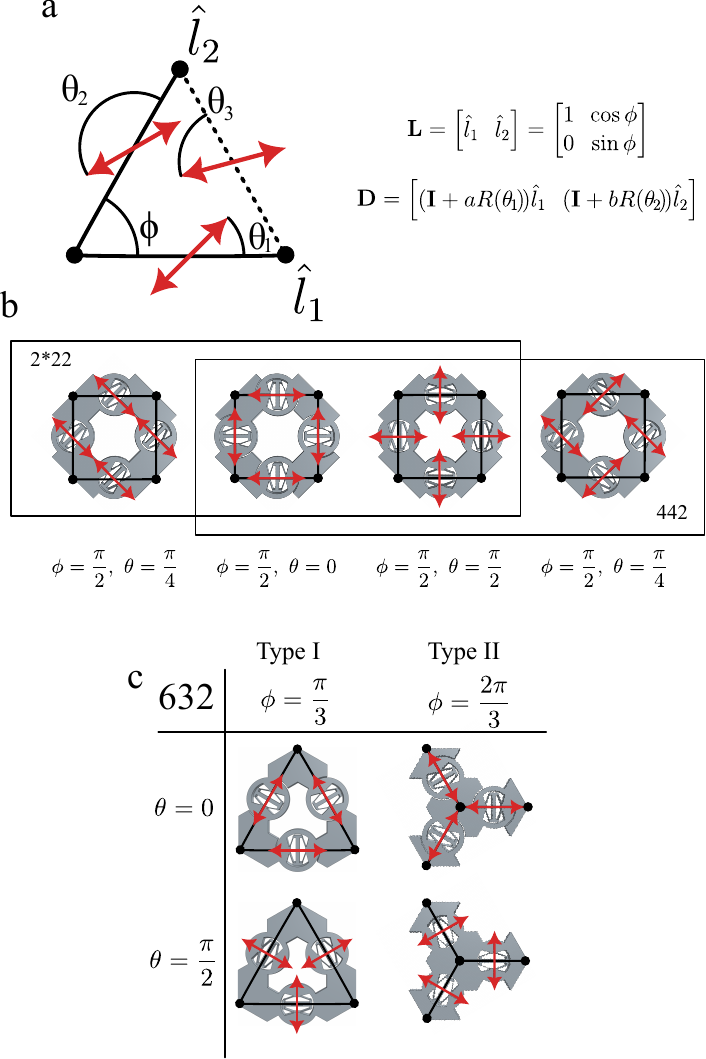}
    \caption{The generalized form of constructing SLM lattices. a) The two lattice vectors form matrix \(\mathbf{L}\) and the deformation gradient, parameterized by \(a\) and \(b\) is found as rotational offsets from the lattice vectors. Under symmetry the SLM rotation angle \(\theta\) is assumed to be the same for every SLM.  b) The two square symmetries use \(\phi=\frac{\pi}{2}\) for their lattice construction, but vary in how the symmetry is applied to realize the other SLM orientations. They are identical for \(\theta = {0, \frac{\pi}{2}}\) where they converged into *442 symmetry. c) 632 symmetry uses two different lattice vector sets, \(\phi = {\frac{\pi}{3}, \frac{2\pi}{3}}\), for Type I and Type II cells respectively. Unlike the square symmetry case there is an additional SLM to account for in the deformation gradient.}
    \label{fig:tiling}
\end{figure}

A general method of constructing compliant extremal materials and analytically determining their zero modes is developed and illustrated in~\autoref{fig:tiling} and algebraically shown in the supplementary file \texttt{extremal\_algebra.m}.
We assume periodic tiling using the lattice vectors \(\mathbf{L} = \begin{bmatrix} \hat{l}_1 & \hat{l}_1 \\ \end{bmatrix}\).
We are free to choose the orientation of our coordinate system and make \(\hat{l}_1 = [1, 0]^T\) for simplicity.
All other vectors (lattice and deformation) can be derived as various rotations of this vector.
We find the Cartesian deformation gradient \(\mathbf{F}\) for each unit cell and then can analyze unique deformation modes as well as the linear strain zero modes.

The generalized lattice vectors are defined with \( \hat{l}_1 \) oriented along the x-axis. 
The second lattice vector \( \hat{l}_2 \) is generated by rotating \( \hat{l}_1 \) by \( \phi \) radians
\begin{equation*}
\hat{l}_2
=
\mathbf{R}(\phi) \hat{l}_2
=
\begin{bmatrix}
    \cos{\phi} \\
    \sin{\phi}
\end{bmatrix}
\end{equation*}
where \(\mathbf{R}(\phi)\) is the standard 2D rotation matrix 
\[
\mathbf{R}(\phi)
=
\begin{bmatrix}
    \cos\phi & -\sin\phi \\
    \sin\phi & \cos\phi
\end{bmatrix}
\]
(\autoref{fig:tiling}a).
Square lattices used in this work have \(\phi = \frac{\pi}{2}\) and hexagonal lattices have \(\phi = \frac{n\pi}{3}\) where \(n=1,2\) (\autoref{fig:tiling}c).
For 632 symmetry, a third vector \( \hat{l}_3 \) must be considered to account for additional DOFs and constraints, which can be incorporated as terms into the first two.

The deformation vector of an SLM is a rotation of the lattice vector it is defined on
\begin{equation*}
    d_i=\mathbf{I} + \xi \mathbf{R}(\theta_i)\hat{l}_i
\end{equation*}
where \(\xi\) represents the free parameters associated with that SLM.
The deformation gradient of an entire unit cell is then 
\begin{equation}
    \mathbf{D}
    =
    \begin{bmatrix}
        d_1 & d_2
    \end{bmatrix}
    =
    \begin{bmatrix}
        (\mathbf{I} + a \mathbf{R}(\theta_1) \hat{l}_1)
        &
        (\mathbf{I} + b \mathbf{R}(\theta_2) \hat{l}_2)
    \end{bmatrix}.
    \label{eq:D}
\end{equation}
The SLM deformation vectors are defined by the individual \(\theta_i\) for \(i=1,2,3\).
For \(\theta_i=0\) the SLM DOF is aligned with the lattice vector it is attached to.
The allows us to again use a rotation of the lattice vector to define the deformation vectors \(d_1\text{ and }d_2\).

The deformation vectors have the free parameters \( a \) and \( b \) for square symmetry as well as \( c \) in 632 symmetry.
These parameters are assumed to lie within the interval [0, 1] and represent the length scale of displacement of the DOF.
The unit cell deformation gradient must simultaneously satisfy all individual SLM DOFS and constraints.

The deformation gradient \(\mathbf{D}\) expresses global deformations in terms lattice vector lines, which may not necessarily be the global Cartesian space.
In general, global deformation gradient can be found from the lattice deformation gradient
\begin{equation*}
\mathbf{F} = \mathbf{L} \mathbf{F}' \mathbf{L}^{-1}    
\end{equation*}
where \(\mathbf{F}'\) is the deformation gradient of the lattice in lattice coordinates and the lattice coordinates \(\mathbf{L}\) work as transformation matrices between the two.
\(\mathbf{D}\) already expresses global deformation in lattice coordinates \(\mathbf{D}=\mathbf{L}\mathbf{F}'\), so to find the Cartesian deformation gradient we use \begin{equation} \mathbf{F} = \mathbf{D} \mathbf{L}^{-1}. \end{equation}
This approach allows us to articulate deformations in a universal coordinate system, thereby generalizing the model to accommodate various lattice symmetries and configurations.

\subsection*{Applying Symmetry}

\begin{table}[tbp]
\centering
\caption{Symmetry enforces different constraints on the lattice shown here for the four types of symmetry explored in this work. \(\theta_1\) is the driven design parameter that defines all other SLM orientations. Square symmetry was restricted to tiling only two SLMs, making \(\theta_3\) undefined for 2*22 and 442 symmetry.}
\label{tab:symmetry_constraints}
\begin{tabular}{lcccc}
\hline
             & 2*22              & 442               & 632 - Type I      & 632 - Type II      \\ \hline
\(\phi\)     & \(\frac{\pi}{2}\) & \(\frac{\pi}{2}\) & \(\frac{\pi}{3}\) & \(\frac{2\pi}{3}\) \\
\(\theta_2\) & -\(\theta_1\)     & \(\theta_1\)      & \(\theta_1\)      & \(\theta_1\)       \\
\(\theta_3\) & -                 & -                 & \(\theta_1\)      & \(\theta_1\)       \\ \hline
\end{tabular}
\end{table}

Applying symmetry introduces constraints to the generalized lattice and deformation construction seen in~\autoref{tab:symmetry_constraints}.
Pure rotational symmetry uses the lattice construction with the constraint \(\theta_i=\theta_j,~\forall i,j\).
Mirror symmetry has the constraint of \(\theta_2 = -\theta_1\), but for mirror symmetry there is no way to define \(\theta_3\) and be mirror symmetric with both \(\theta_1\) and \(\theta_2\).
This means that the 632 wallpaper group can not, in general, produce mirror symmetric deformation gradients across \(\theta\), which is observed on the extremal gamut.
The trajectory of an isotropic extremal material from one state to another passes through the anisotropic regime, implying that isotropy is not allowed while maintaining symmetry.

\subsection*{Analyzing the Deformation Gradient}
\begin{figure}[htbp]
    \centering
    \includegraphics{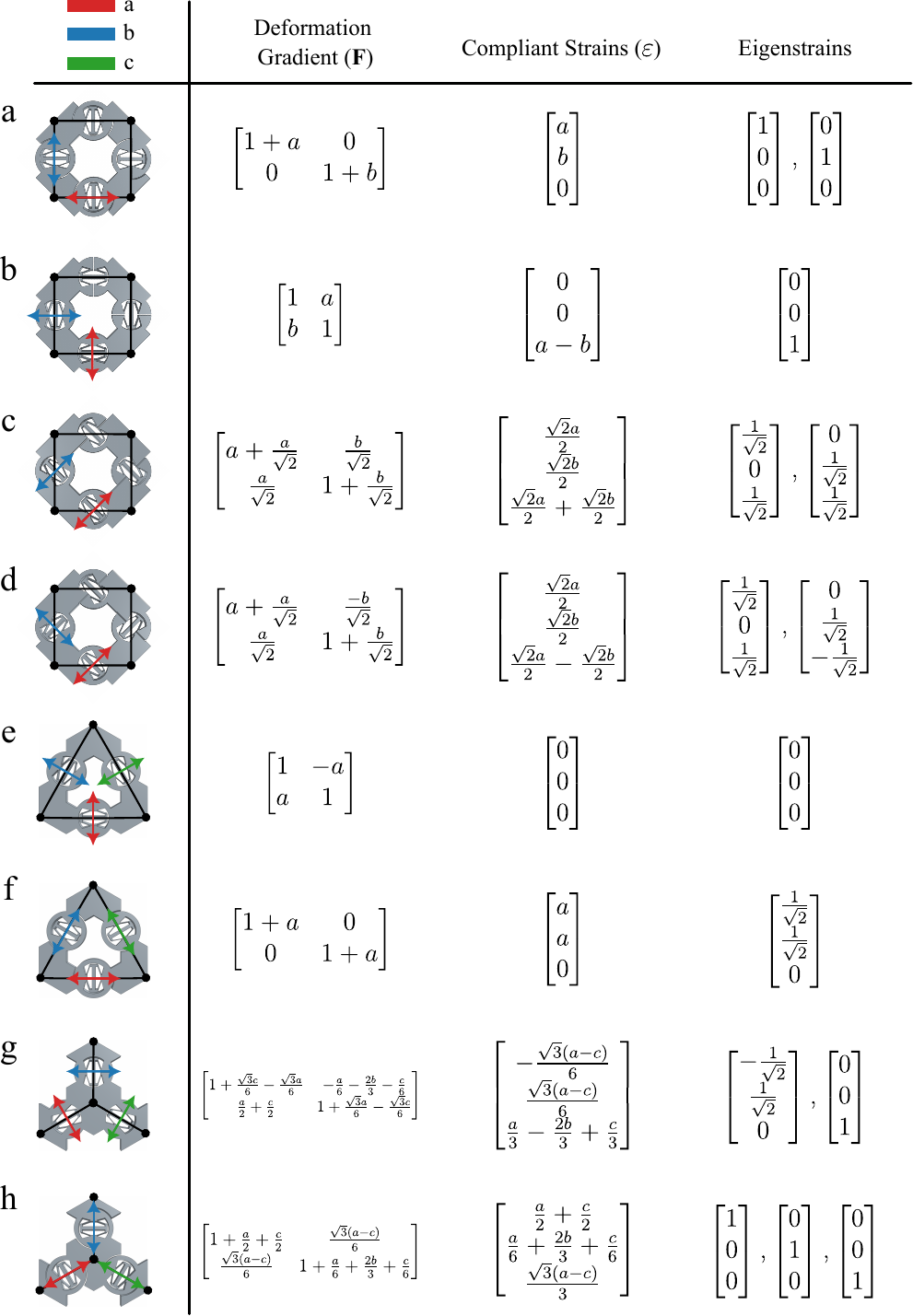}
    \caption{The three symmetry groups examined in this study are presented at specific angles in~\autoref{tab:symmetry_constraints}. Deformation vectors are indicated by colored arrows, with a legend above explaining the correspondence of each vector to the variables a, b, or c. Deformation gradients are calculated by fulfilling all independent SLMs within the cell concurrently. Compliant strains are determined as linear strains and are displayed in vector form. These strains can also be expressed as linearly independent eigenstrains, which are the unique strains compliant with a material constructed with the specified DOF.}
    \label{fig:big_one}
\end{figure}

We can construct an idealized deformation gradient for the unit cells by linearly combining multiple SLM deformation gradients (\autoref{eq:F}) and resolving any constraints. 
This is shown in \autoref{fig:big_one} and derived in \texttt{extremal\_algebra.m}, where the deformation gradient is derived from the generalized lattice formulation in each case.  
There are two independent SLMs in square symmetry (red and blue) and three independent SLMs for hexagonal symmetry (red, blue, and green). 
These are color coded to highlight how each SLM is allowed to independently deform and illuminate the relationship with the deformation gradient.

For example, \autoref{fig:big_one}a is the 2*22 \(\theta = 0\) bimodal material. 
This material allows horizontal deformation along the horizontal lattice lines and vertical deformation along vertical lattice lines.
This is shown as the independent red and blue arrows, but also conveyed in the deformation gradient with the independent parameters \(a\) and \(b\) each allowed to modify the deformation gradient.
The Type I hexagonal materials have a third SLM which operates as a constraint on the unit cell by ensuring \(|\hat{l}_1|=|\hat{l}_2|\). 
Thus, the deformation gradient actually loses free parameters, seen in \autoref{fig:big_one}e and f having only the free parameter \(a=b\).
Meaning, that if the SLM that corresponds to \(a\) deforms then then SLM corresponding to \(b\) must also deform.

Analyzing the deformation gradient directly can provide insight into the material that is lost in linear elasticity.
For example, the isotropic nullmode material (\autoref{fig:big_one}e) has deformation gradient 
\[
\begin{bmatrix}
    1 & -a \\
    a & 1
\end{bmatrix}
\]
indicating rotation that is possible in the unit cell.
The eigenvectors (\(Q\)) and eigenvalues (\(\Lambda\)) of this material's deformation gradient are
\[
Q
=
\begin{bmatrix}
    -i & i \\ 
    1 & 1
\end{bmatrix}
,\quad
\Lambda
=
\begin{bmatrix}
    1 - ai & 0 \\
    0 & 1 + ai
\end{bmatrix}
\]
confirming that rotational deformation of the material is possible.
However, this information is lost in the translation to Cauchy elasticity which does not account for rotations.
This is seen by the compliant strain equaling the zero vector in \autoref{fig:big_one}, seemingly indicating that no deformation is allowed.
The use of polar DOFs is of substantial interest~\cite{Frenzel2017-of, Nassar2019-wy, Xu2020-mc} and an examination using micropolar or Cosserat elasticity theories would be required to explore this concept further.

The compliant linear strain provides key insight into a material's behavior by revealing the zero modes of a linear elastic material.
We find the zero modes by converting the deformation gradient to strain which, in the limit of infinite compliance, satisfies
\begin{equation}
    \sigma=0 = C \varepsilon
    \label{eq:hooke_zero}
\end{equation}
implying that the compliant strains form the nullspace of \(C\).
In reality, the elasticity matrix \(C\) is always symmetric positive definite, meaning that \autoref{eq:hooke_zero} can only be satisfied when \(\varepsilon = [0,0,0]^T\).
However, if the compliance is weak enough it will still approximately satisfy \autoref{eq:hooke_zero}.
Equivalently, the compliant strains are the eigenvectors corresponding to zero-value eigenvalues, meaning that the number of independent zero-energy strain vectors (i.e., the nullity of \(C\)) are the extremal mode of the material.
This is seen in \autoref{fig:big_one}, where nullmode materials have no eigenstrains (outside the trivial zero vector), unimodal materials have one eigenstrain, bimodal materials have two, and the trimodal materials have three.

\edit{
\section{An Analytical Model for the Compliance Matrix \(S\)}
\subsection*{Derivation}
The analytical compliance matrix can be developed using the same linear strain defined in~\autoref{eq:strain}.
We express the Voigt strain \(\varepsilon=[\varepsilon_{xx}, \varepsilon_{yy},\gamma_{xy}]^T\) in terms of the strain-DOF operator matrix \(B(\theta)\) and the DOFs \(q=[a,b]^T\) where \(a\) and \(b\) are the DOFs for the individual SLMs shown in~\autoref{eq:D} as
\begin{equation}
    \varepsilon(\theta)=B(\theta)q.
\end{equation}
\(B(\theta)\) then dictates the allowable deformation modes of the lattice irrespective of the actual SLM deformation length scales.
We then model each SLM stiffness with an axial spring \(k_i\), letting \(K=\text{diag}(k_a,k_b)\).
Minimizing the internal energy \(U=\frac{1}{2}q^TKq\) subject to the imposed strain leads to \(q=K^{-1}B^T\sigma\) and therefore
\begin{equation}
    S(\theta)=B(\theta)K^{-1}B(\theta)^T
    \label{eq:S}
\end{equation}
which is the effective compliance matrix seen at the macroscale.
If all the SLMs are equal stiffness (\(k_i=k\)) \autoref{eq:S} reduces to \(S(\theta)=\frac{1}{k}B(\theta)B(\theta)^T\).
}

\edit{
\subsection*{Fitting Analytical Models to Data}
Analytical models were fit to both the measured data and the FEA square-symmetric (2*22 and 442) simulated data for direct comparison.
This provides a robust and comprehensive method of directly comparing the material characteristics and behavior across \(\theta\) rather than on a point-by-point basis.
First, the analytical models were derived for the 2*22 and 442 materials in the lattice orientation used for testing and simulation. 
Assuming \(k_i=k=1\), the characteristic behaviors are then
\begin{equation}
    S(\theta)_{2*22} = 
    \begin{bmatrix}
        \frac{1}{2}\left(1+\sin(2\theta)\right) & \frac{1}{2}\cos(2\theta) & 0 \\
        \frac{1}{2}\cos(2\theta) & \frac{1}{2}(1-\sin(2\theta)) & 0 \\
        0 & 0 & \cos(2\theta) + 1
    \end{bmatrix}
    \label{eq:S222}
\end{equation}
and
\begin{equation}
    S(\theta)_{442} = 
    \begin{bmatrix}
        \frac{1}{2} & \frac{1}{2}\cos(2\theta) & -\frac{1}{2}\sin(2\theta) \\
        \frac{1}{2}\cos(2\theta) & \frac{1}{2} & \frac{1}{2}\sin(2\theta) \\
        -\frac{1}{2}\sin(2\theta) & \frac{1}{2}\sin(2\theta) & \cos(2\theta) + 1
    \end{bmatrix}.
    \label{eq:S442}
\end{equation}
We generalize the models to optimize over a set of linear coefficients \(C\)
\begin{equation}
    S(\theta)_{2*22}^{opt} = 
    \begin{bmatrix}
        c_1\sin(2\theta)+c_2 & c_3\cos(2\theta)+c_4 & 0 \\
        c_3\cos(2\theta)+c_4 & c_2-c_1\sin(2\theta) & 0 \\
        0 & 0 & c_5\cos(2\theta) + c_6
    \end{bmatrix}
\end{equation}
and
\begin{equation}
    S(\theta)_{442}^{opt} = 
    \begin{bmatrix}
        c_1 & c_2\cos(2\theta)+c_3 & c_4\sin(2\theta) \\
        c_2\cos(2\theta) + c_3 & c_1 & -c_4\sin(2\theta) \\
        c_4\sin(2\theta) & -c_4\sin(2\theta) & c_5\cos(2\theta) + c_6
    \end{bmatrix}
\end{equation}
optimizing over all \(c_i \in C\) for each material set independently.
The optimization function then becomes
\begin{equation}
    \begin{aligned}
        \min_{C} \quad & \sum_{\theta_i \in \Theta} \frac{(S(\theta_i;C)^{\text{opt}} - \hat{S}(\theta_i))^2}{(\hat{S}(\theta_i))^2} \\
        \text{s.t.} \quad & \det(S(\phi)) > 0, \quad \phi = \{0^\circ, 1^\circ, 2^\circ, \ldots, 180^\circ\}
    \end{aligned}
\end{equation}
where \(\hat{S}(\theta_i)\) represents either the measured data or the FEA modeled data for each independent model fit respectively, \(\Theta\) represents the set of angles that resolve the data (0\textdegree{} to 180\textdegree{} in 22.5\textdegree{} increments for measured data and 5\textdegree{} increments for FEA data) and the constraint on the determinant is applied to a more fine angular increment \(\phi\) every 1\textdegree{} to the model to ensure the it is physically realistic at all points.
The final optimized coefficients for the modeled data are provided in \autoref{tab:fit-coefs}.
}

\begin{table}[]
\captionsetup{labelfont={color=black}, textfont={color=black}}
\caption{Analytical model fit coefficients for each measured and FEA data set. Reference \autoref{eq:S222} and \autoref{eq:S442} for placement of coefficients in \(S\) matrices.}
\label{tab:fit-coefs}
\centering
{\color{black}
\begin{tabular}{@{}rcccccc@{}}
\toprule
Data Set      & $c_1$ & $c_2$ & $c_3$  & $c_4$  & $c_5$ & $c_6$ \\ \midrule
2*22 Measured & 2.606 & 3.018 & 2.336  & 0.072  & 3.720 & 4.908 \\
2*22 FEA      & 3.092 & 3.150 & 3.061  & -0.003 & 6.197 & 6.321 \\
442 Measured   & 2.868 & 2.591 & -0.120 & -2.346 & 2.753 & 3.727 \\
442 FEA       & 2.904 & 2.826 & 0.064  & -2.855 & 5.636 & 5.755 \\ \bottomrule
\end{tabular}
}
\end{table}

\begin{figure}
    \centering
    \includegraphics[width=0.5\linewidth]{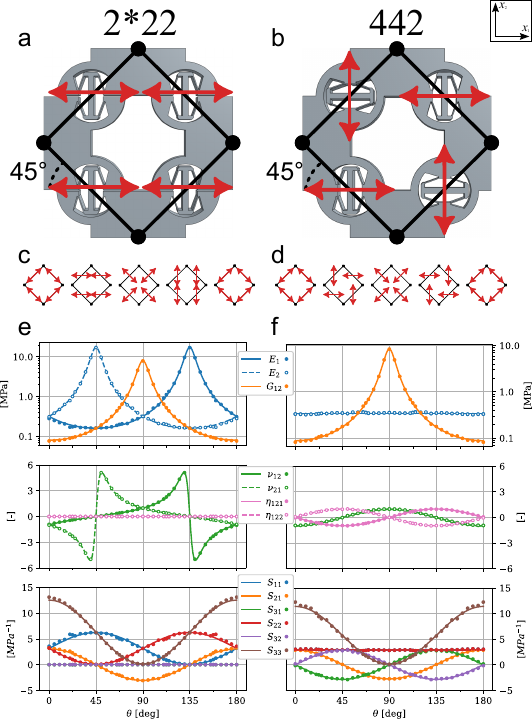}
    \caption{\edit{A twin figure to the main text Figure 2, here showing FEA simulated and FEA fit model data. CAD models used for FEA for 2*22 (a) and 442 (b) square symmetry materials shown at a \(\theta=45\degree\) configuration. The illustrations below each image (c and d) show the DOF alignment for each symmetry as a function of \(\theta\) and correspond to the major x-axis grid lines in the plots below. The data shown in (e) and (f) show experimental data (dots) and corresponding analytical fits (solid/dashed lines). From top to bottom: the directional Young's and shear moduli ($E_1,E_2$ and $G_{12}$); the Poisson's and normal-shear coupling ratios ($\nu_{12},\nu_{21}$ and $\eta_{121},\eta_{122}$); and finally the individual $S$ matrix components which were used to fit the model directly. Compared to the main text Figure 2, interquartile ranges are not shown due to each data point having only one simulation run.}}
    \label{fig:fig2twin}
\end{figure}

\edit{
The measured and modeled data are presented in the main text Figure 2. Here we include a twin figure (\autoref{fig:fig2twin} to show the FEA simulated and FEA fit model.
Here we see even better agreement between the FEA data and model as compared to the measured data and its respective model. 
Notably, the FEA model predicts a higher stiffness than the measured results indicate; which is also borne out in the larger coefficients of \autoref{tab:fit-coefs}. 
However, the general trends are all qualitatively well-matched (e.g., 2*22 peaks in moduli or asymptotes in Poisson's ratio; 442 constant Young's moduli sinusoidal normal-shear coupling).
This then validates the models derived in \autoref{eq:S222} and \autoref{eq:S442} as characteristic of the materials' behavior across the full range of \(\theta\) and indicates suitability for use against the measured data in the main text.
}

\section{Eigenvalue Normalization and Gamut}
To study the transitions in SLM lattices, we introduce the concept of the "extremal material gamut" based on normalized eigenvalues ordered from smallest to largest.
\begin{equation}
    \hat{\lambda}_i = \frac{\lambda_i}{\lambda_3}, \quad \text{for } i \in \{1, 2, 3\} \text{ and } \lambda_1 \leq \lambda_2 \leq \lambda_3
\end{equation}
Normalization of eigenvalues fulfills two functions: it facilitates rapid evaluation of extremal materials by highlighting smaller eigenvalues and permits comparisons across different material symmetries by focusing on extremal attributes instead of direct stiffness.

Furthermore, normalized eigenvalues help define the eigenvalue gamut, a feasible space delineated by \(\hat{\lambda}_1 = \hat{\lambda}_2\), \(\hat{\lambda}_2 = \hat{\lambda}_3 = 1\), and \(\hat{\lambda}_1 = 0\) within the unit square.
Isotropic materials --- characterized by two equal eigenvalues~\cite{Milton1995-rd} --- are located along the diagonal and the upper edge of this triangle.
The left boundary, marking anisotropic limits, exhibits no equal eigenvalues except at the endpoints, but always includes extremal points, with \(\hat{\lambda}_1 = 0\) denoting ideal unimodal and bimodal materials at \((0,1)\) and \((0,0)\) respectively, encompassing all intermediate unimodal points.

\section{Test and Characterization of Array}
\begin{figure}[htbp]
    \centering
    \includegraphics[width=\textwidth]{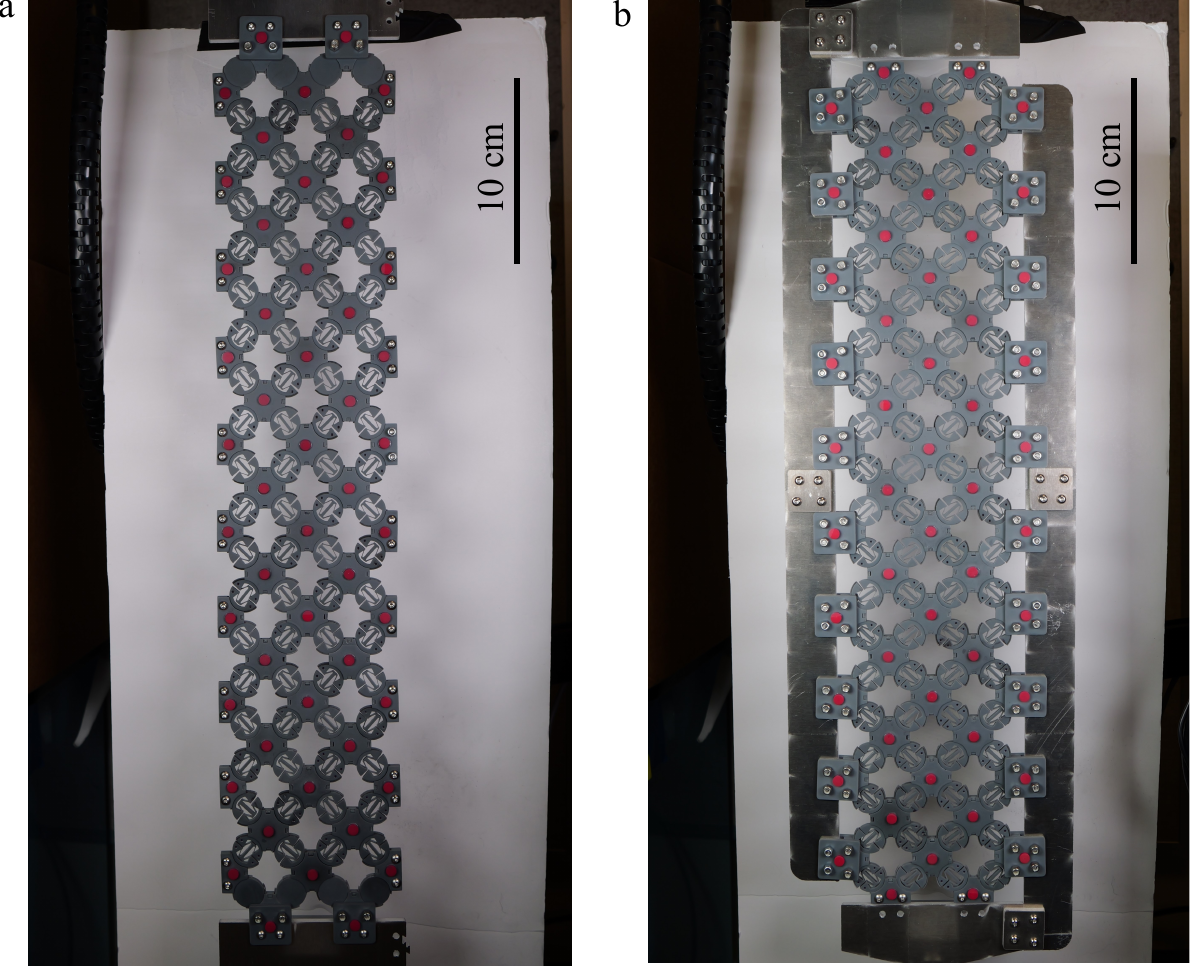}
    \caption{The \(4 \times 18\) array under tensile loading (a) and shear loading (b). Red markers are used for optical tracking to extract local strain information in the array.}
    \label{fig:array}
\end{figure}

\begin{figure}[htbp]
    \centering
    \includegraphics[width=\textwidth]{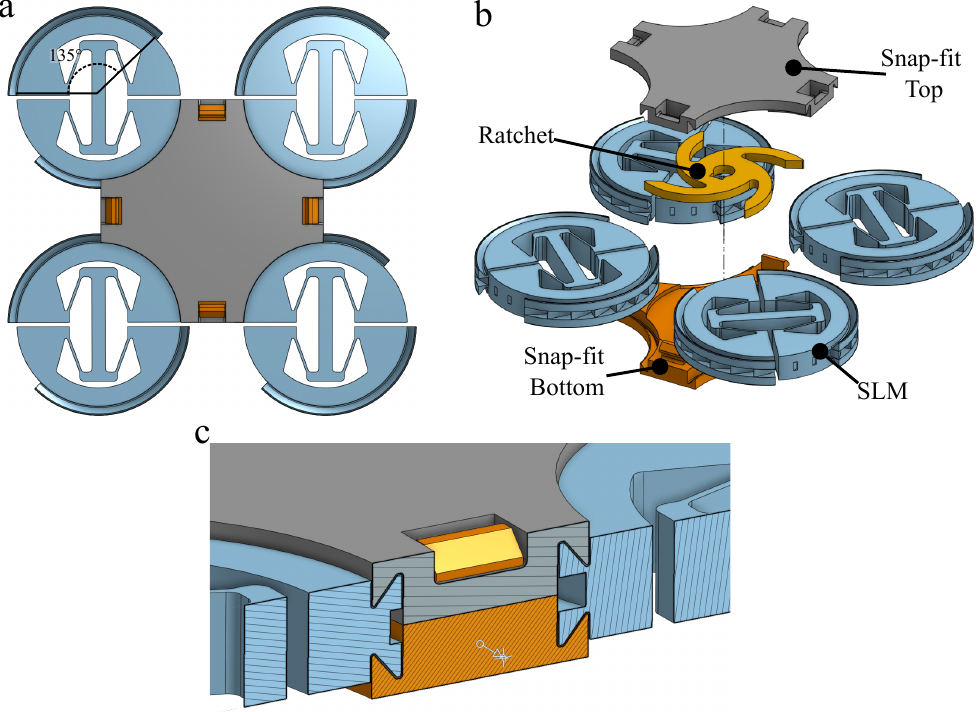}
    \caption{CAD of the SLM reprogrammable array unit cell. SLMs are arranged on a square lattice (a) and attached by running a dovetail through a snap-fit enclosure (b) so that the SLM is captured at least on one side. There is a hidden, internal ratchet spring system to help align and secure the SLMs during testing. (c) shows a close-up section view of the dovetail, illustrating how the SLM is allowed to pass through the enclosure, while maintaining contact.}
    \label{fig:cad-array}
\end{figure}

\subsection*{Array Construction}
A \(4 \times 18\) square lattice array was fabricated to validate and characterize the 2*22 and 442 materials (\autoref{fig:array}). The design incorporates snap-fit connectors and a rotational dovetail, facilitating reconfigurable and easy assembly (\autoref{fig:cad-array}). 
This modular construction allows for the easy replacement of parts because no parts are rigidly connected, instead relying on friction and other forces to maintain contact. 
A ratcheting scheme aligns and secures the SLMs during testing (\autoref{fig:cad-array}b).
The close fit of the dovetail seen in the section view \autoref{fig:cad-array}c enables the SLM to pivot while maintaining its connection during loading for a range of \(\theta\). 

The dovetail's length determines the array's available configuration space. 
Dovetails in the samples extend over \(135^\circ\) (\autoref{fig:cad-array}a), allowing testing within \(\theta \in [0,135^\circ]\). 
However, this precludes reprogrammability within \(\theta \in (135^\circ,180^\circ)\), where the dovetails contact the snap-fit base, immobilizing the SLM and eliminating all compliance. 
This restriction serves as a tunable parameter, offering potential areas for future investigation.

\subsection*{Test, Measurement, and Characterization}
The array in \autoref{fig:array} was tested on a universal testing machine (UTM) (Instron 5965) in a quasi-static manner.
Two symmetry patterns (2*22 and 442) were tested, in two different loading conditions (tensile and shear), with approximately five tests per \(\theta\) value for \(\theta \in {0, 22.5^\circ, 45^\circ, 67.6^\circ, 90^\circ, 112.5^\circ, 135^\circ}\).
The force vs. displacement curves and videos of the array deforming array were recorded for testing at various \(\theta\) configurations.
The global force was used for characterization, while the middle third of the optical trackers were used to capture deformation behavior.
A subset of the trackers were used to avoid boundary effects at the top and bottom end of the array.

\section{Table of Material Values}
Material values for the materials seen in Figure 1 from the main document are shown in \autoref{tab:2star22-matprop} and \autoref{tab:632-matprop}.

\begin{sidewaystable}[h]
\caption{Selection of 2*22 Simulated Material Properties}
\label{tab:2star22-matprop}
\begin{tabular}{lllllllllllll}
\hline
\multicolumn{1}{c}{\(\theta\)} &
  \multicolumn{1}{c}{mode} &
  \multicolumn{1}{c}{\(E_1\) [MPa]} &
  \multicolumn{1}{c}{\(E_2\) [MPa]} &
  \multicolumn{1}{c}{\(G\) [MPa]} &
  \multicolumn{1}{c}{\(\nu_{12}\)} &
  \multicolumn{1}{c}{\(\eta\)} &
  \multicolumn{1}{c}{\(\lambda_1\) [MPa]} &
  \multicolumn{1}{c}{\(\lambda_2\) [MPa]} &
  \multicolumn{1}{c}{\(\lambda_3\) [MPa]} &
  \multicolumn{1}{c}{\(\hat\lambda_1\)} &
  \multicolumn{1}{c}{\(\hat\lambda_2\)} &
  \multicolumn{1}{c}{\(\hat\lambda_3\)} \\ \hline
0\textdegree  & m=2 & 0.280 & 0.310  & 0.076 & -0.909 & 0.001  & 0.150 & 0.153  & 6.740  & 0.022 & 0.023 & 1.000 \\
45\textdegree & m=2 & 0.162 & 17.099 & 0.160 & 0.001  & -0.001 & 0.162 & 0.320  & 17.100 & 0.009 & 0.019 & 1.000 \\
90\textdegree & m=1 & 0.317 & 0.317  & 7.880 & 0.980  & 0.000  & 0.160 & 15.760 & 16.040 & 0.010 & 0.983 & 1.000 \\ \hline
\end{tabular}
\end{sidewaystable}

\begin{sidewaystable}[h]
\caption{632 Simulated Material Properties}
\label{tab:632-matprop}
\begin{tabular}{lllllllllllll}
\hline
\(\theta\) &
  \multicolumn{1}{c}{mode} &
  \multicolumn{1}{c}{\(E\) [MPa]} &
  \multicolumn{1}{c}{\(G\) [MPa]} &
  \multicolumn{1}{c}{\(K\) [MPa]} &
  \multicolumn{1}{c}{\(K/G\)} &
  \multicolumn{1}{c}{\(\nu\)} &
  \multicolumn{1}{c}{\(\lambda_1\) [MPa]} &
  \multicolumn{1}{c}{\(\lambda_2\) [MPa]} &
  \multicolumn{1}{c}{\(\lambda_3\) [MPa]} &
  \multicolumn{1}{c}{\(\hat\lambda_1\)} &
  \multicolumn{1}{c}{\(\hat\lambda_2\)} &
  \multicolumn{1}{c}{\(\hat\lambda_3\)} \\ \hline
90\textdegree & m=0 & 16.000 & 6.040 & 12.000 & 1.987  & 0.330  & 12.030 & 12.080 & 24.070 & 0.500 & 0.502 & 1.000 \\
0\textdegree  & m=1 & 0.526  & 2.190 & 0.140  & 0.064  & -0.880 & 0.280  & 4.378  & 4.382  & 0.064 & 0.999 & 1.000 \\
90\textdegree & m=2 & 0.394  & 0.102 & 6.680  & 65.490 & 0.970  & 0.200  & 0.204  & 13.360 & 0.015 & 0.015 & 1.000 \\
0\textdegree  & m=3 & 0.131  & 0.097 & 0.050  & 0.513  & -0.320 & 0.099  & 0.193  & 0.195  & 0.510 & 0.993 & 1.000 \\ \hline
\end{tabular}
\end{sidewaystable}

\section*{Media Legends}

\subsection{Movie M1:} In-plane, orientation dependent material properties of the 2*22 array. As the angle of the Straight Line Mechanisms (SLM) vary relative to the C4 symmetry of the lattice, the materials relative Young's Modulus and shear modulus evolve. The material transitions from positive Poisson Ratio, to zero and ultimately a negative Poisson's ratio material.

\subsection{Movie M2:} In-plane, orientation dependent material properties of the 442 array. As the angle of the Straight Line Mechanisms (SLM) vary relative to the C4 symmetry of the lattice, the materials relative Young's Modulus and shear modulus evolve by clocking and changing magnitude.

\subsection{Movie M3:} In-Situ array reprogramming. We demonstrate the process of changing the arrays properties without reconstruction of the array.

\bibliographystyle{MSP}
\bibliography{bib.bib}